\documentclass[sigconf, nonacm]{acmart}
\AtBeginDocument{%
  }

\acmISBN{978-1-4503-XXXX-X/2018/06}

\begin{document}

\title{Connecting Feedback to Choice: Understanding Educator Preferences in GenAI vs. Human-Created Lesson Plans in K-12 Education -- A Comparative Analysis}

\author{Shawon Sarkar}
\authornote{Corresponding author}
\email{ss288@uw.edu}
\affiliation{%
  \institution{University of Washington}
  \city{Seattle}
  \state{WA}
  \country{USA}
}

\author{Min Sun}
\email{misun@uw.edu}
\affiliation{%
  \institution{University of Washington}
  \city{Seattle}
  \state{WA}
  \country{USA}
}

\author{Alex Liu}
\email{alexliux@uw.edu}
\affiliation{%
  \institution{University of Washington}
  \city{Seattle}
  \state{WA}
  \country{USA}
}

\author{Zewei Tian}
\email{ztian27@uw.edu}
\affiliation{%
  \institution{University of Washington}
  \city{Seattle}
  \state{WA}
  \country{USA}
}

\author{Lief Esbenshade}
\email{lief@uw.edu}
\affiliation{%
  \institution{University of Washington}
  \city{Seattle}
  \state{WA}
  \country{USA}
}

\author{Jian He}
\email{kevin@hensuninnovation.com}
\affiliation{%
  \institution{Hensun Innovation LLC}
  \city{Seattle}
  \state{WA}
  \country{USA}
}

\author{Zachary Zhang}
\email{zac@hensuninnovation.com}
\affiliation{%
  \institution{Hensun Innovation LLC}
  \city{Seattle}
  \state{WA}
  \country{USA}
}

\thanks{Parts of this study and its findings were presented at the Society for Research on Educational Effectiveness (SREE) Annual Conference, 2024.}

\renewcommand{\shortauthors}{Sarkar et al.}

\begin{abstract}
As generative AI (GenAI) models are increasingly explored for educational applications, understanding educator preferences for AI-generated lesson plans is critical for their effective integration into K-12 instruction. This exploratory study compares lesson plans authored by human curriculum designers, a fine-tuned LLaMA-2-13b model trained on K-12 content, and a customized GPT-4 model to evaluate their pedagogical quality across multiple instructional measures: warm-up activities, main tasks, cool-down activities, and overall quality. Using a large-scale preference study with K-12 math educators, we examine how preferences vary across grade levels and instructional components. We employ both qualitative and quantitative analyses. The raw preference results indicate that human-authored lesson plans are generally favored, particularly for elementary education, where educators emphasize student engagement, scaffolding, and collaborative learning. However, AI-generated models demonstrate increasing competitiveness in cool-down tasks and structured learning activities, particularly in high school settings. Beyond quantitative results, we conduct thematic analysis using LDA and manual coding to identify key factors influencing educator preferences. Educators value human-authored plans for their nuanced differentiation, real-world contextualization, and student discourse facilitation. Meanwhile, AI-generated lesson plans are often praised for their structure and adaptability for specific instructional tasks. Findings suggest a human-AI collaborative approach to lesson planning, where GenAI can serve as an assistive tool rather than a replacement for educator expertise in lesson planning. This study contributes to the growing discourse on responsible AI integration in education, highlighting both opportunities and challenges in leveraging GenAI for curriculum development.
\end{abstract}



\maketitle

\section{Introduction}
Developing rigorous, engaging, and personalized lesson plans is a cornerstone of effective K-12 education, yet it remains a time-consuming and cognitively demanding task for educators. According to the TALIS 2018 Survey \cite{/content/publication/1d0bc92a-en}, U.S. educators spend an average of seven hours per week on lesson planning, with an additional three hours required for students with diverse needs, such as disabilities or language barriers. This workload detracts from direct student interaction—arguably the most rewarding aspect of teaching.

Recent advancements in Artificial Intelligence (AI), particularly generative AI (GenAI) powered by large language models (LLMs), offer promising solutions to streamline lesson planning. A May 2024 Impact Research survey \cite{walton2024value} found that 79\% of teachers reported using GenAI tools like ChatGPT \cite{ChatGPT2024} for lesson planning. Additionally, specialized educational AI platforms (e.g., MagicSchool, School AI) are rapidly emerging, further integrating AI into K-12 education \cite{mollick2023using, kasneci2023chatgpt}. Despite this growing adoption, GenAI tools still exhibit critical limitations: their lack of domain-specific knowledge often results in lesson plans that are contextually inappropriate, pedagogically unsound, or factually inaccurate \cite{brown2020language, radford2019language}. Educators report challenges such as insufficient depth in subject concepts, mathematical errors, and poor understanding of pedagogy and student learning progression.

While prior research has explored the capabilities of LLMs in general language tasks \cite{bommasani2021opportunities, hendrycks2021measuring}, empirical studies directly comparing AI-generated and human-created lesson plans remain scarce. This gap leaves educators and institutions without concrete evidence to guide AI adoption in educational content creation. Our study addresses this by systematically comparing GenAI-generated lesson plans with those crafted by experienced curriculum designers in a K-12 setting.

To evaluate the viability of LLM-generated lesson plans, we conducted a comparative study using pairwise evaluation methods. We assessed lesson plans generated by models such as customized prompt-based GPT-4 and fine-tuned LLaMA-2-13b against human-created plans. Educators rated these plans across four key dimensions: relevance, quality, and usefulness of the warm-up, main task (explain and reinforce), and cool-down sections, along with overall quality. By employing pairwise comparisons with these custom evaluation measures, we captured nuanced educator preferences that traditional evaluation metrics might overlook. Our mixed-method analysis quantitatively assessed GenAI models' performance and qualitatively examined user feedback, identifying AI-generated content's strengths and weaknesses.

Thus this study aims to (i) evaluate GenAI models' lesson planning capabilities against those of professional curriculum designers across different grade levels, and (ii) identify strengths and limitations to inform the development of more effective, domain-specific GenAI tools for education.

Our findings suggest that, even with minimal domain specialization -- using fine-tuned zero-shot prompts and labeled educational data -- GenAI models can produce lesson plans that rival, or in some cases exceed, those created by human experts. The results highlight the potential of domain-specific tuning in GenAI tools to enhance pedagogical quality while reducing educators' workloads, contributing to the broader discourse on AI's role in education.

\section{Background}
This section reviews the existing literature on lesson planning, AI applications in education, and the motivation for this study.

\subsection{Lesson Planning and Its Importance in Education}
Lesson planning is fundamental to effective teaching, providing a structured framework or roadmap for instruction, student engagement, and assessment. For example, a well-designed mathematics lesson plan may consists of four key sections or learning stages: a\textit{ warm up}, \textit{main tasks} (\textit{explain} and \textit{reinforce}), and a \textit{cool down}. Warm up activities review and connect with students' prior knowledge, and stimulate students' interest in the new mathematical concept. The explain stage involves guided instruction, modeling, and interactive discussions. The reinforce phase includes individual practice, collaborative exercises, and feedback opportunities. Finally, the cool-down section summarizes key learnings and connects to future lessons \cite{choppin2022role}.

High-quality lesson planning ensures alignment with educational standards (e.g., Common Core Learning Standards \cite{CCSS2024}) and directly impacts student learning outcomes \cite{farrell2002lesson, john2006lesson, choppin2022role}. However, teachers -- especially early-career educators -- face considerable challenges in designing lesson plans that balance content rigor, pedagogical effectiveness, and classroom adaptability \cite{jerrim2019teaching, jones2022teachers, jerrim2021high, sun2018black, reeves2017influence}. The cognitive demands of lesson planning are further amplified by the need to accommodate diverse student populations, including those with linguistic and special education needs \cite{irwin2023report, kaufman2018changes, hertel2017examining, polikoff2021beyond}.

As educators juggle multiple responsibilities, the time and effort required to create detailed lesson plans can be overwhelming. To mitigate these challenges, educators often turn to Open Educational Resources (OERs) (e.g., Achieve the Core, BetterLesson, Gooru, OER Commons) \cite{achievethecore, betterlesson, gooru, illmath, OERCommons, sharemylesson, tpt}. However, OERs primarily provide static content rather than personalized, pedagogically grounded lesson plans \cite{tong2023investigating, hertel2017examining}. These challenges have led to a growing interest in tools and technologies that can assist with or even automate parts of the lesson planning process, helping to reduce educators' workloads while maintaining high educational standards.

\subsection{AI and LLM Applications in Education: Capabilities and Limitations}
The rise of LLMs such as GPT-4, Claude, Gemini, and LLaMA \cite{brown2020language, anthropic2024claude, team2024gemini, touvron2023LLaMA} has accelerated AI's integration into education \cite{diliberti2024using}. These models show promise in automating lesson planning, content generation, student assessment, and personalized learning \cite{holmes2019artificial}. AI tools can dynamically generate instructional materials across subjects and grade levels, potentially reducing educators' workload while maintaining content quality \cite{pedro2019artificial, wang2024artificial, weng2024integrating}.

Despite these advantages, significant challenges remain. General-purpose LLMs lack domain-specific pedagogical knowledge, leading to content that may be factually incorrect, pedagogically misaligned, or culturally insensitive \cite{bender2021dangers, bommasani2021opportunities, eloundou2023gpts, kasneci2023chatgpt, marcus2020next}. The unpredictability of AI-generated lesson plans raises concerns about their reliability, relevance, and adherence to educational standards \cite{weidinger2021ethical}. Furthermore, educators report issues such as insufficient depth in subject matter, lack of contextual specificity, and frequent errors in STEM-related content \cite{kasneci2023chatgpt, marcus2020next}. Therefore, while LLMs can assist in generating content, their outputs must be rigorously evaluated to ensure pedagogical soundness and contextual relevance.

Emerging GenAI-based lesson planning platforms\citep[e.g.][]{magicschool, teachologyai, lessonplansai}, offers to address these issues by customizing LLMs and offering different features designed for educational contexts. However, their effectiveness remains largely unverified. Many of these tools still lack rigorous validation against educational research and pedagogical best practices, limiting their adoption in formal K-12 settings.

The current state of AI-based lesson planning tools presents both opportunities and risks, and further empirical studies are required to evaluate their effectiveness in real-world educational environments \cite{wang2024artificial, lin2021engaging}. As an early effort, in this study, we contribute by examining user preferences between professional human- and AI-generated lesson plans, offering insights into these approaches' comparative strengths and limitations.

\section{User Study Setup and Data Collection}
To rigorously evaluate the effectiveness of AI-generated lesson plans compared to those created by human curriculum designers, we conducted a controlled user study with experienced K-12 educators. Participants assessed lesson plans generated by three different sources: a customized prompt-based GPT-4 model, a fine-tuned LLaMA-2-13b model, and human curriculum designers. The study examined educators' preferences across multiple dimensions to provide insights into the comparative strengths and weaknesses of AI-assisted lesson planning in K-12 education.

\subsection{Study Instruments and System Setup}
To ensure a robust and fair evaluation process, we utilized the following instruments:
\begin{itemize}
    \item \textbf{Query set:} We created a pool of 400 unique queries, each comprising a subject, grade level (K-12), a Common Core State Standard for Mathematics (CCSSM) \cite{CCSS2024}, a lesson plan title ($\le 100$ words), and 2-3 learning objectives. Each query generated three lesson plans -- one by customized GPT-4, one by fine-tuned LLaMA-2-13b, and one retrieved from a open-source lesson plan database. To maintain consistency, we focused exclusively on mathematics lesson plans.
    \item \textbf{Lesson plans:}
    \begin{itemize}
        \item Human curriculum designer-created lesson plans: Retrieved from a database of lesson plans sourced from online open-source repositories, such as \textit{Illustrative Mathematics} \cite{illmath} via a semantic search algorithm using \textit{text-embedding-ada-002} \cite{embededdingAda}.
        \item AI-generated lesson plans: The LLaMA-2-13b model \cite{touvron2023LLaMA}, one of the state-of-the-art art open-source models during the time of the user study, was fine-tuned on expert-designed lesson plans, ensuring exposure to high-quality, pedagogically sound content. The GPT-4 \cite{gpt4} model was optimized via prompt engineering, refining its ability to generate structured lesson plans sequentially across key sections (warm-up, explain, reinforce, and cool-down). Both models used identical system prompts, model parameters (e.g., temperature, tokens), and query inputs (subject, grade, title, learning objectives, and CCSSM), ensuring direct comparability with human-created lesson plans.
    \end{itemize}
    Each lesson plan followed a standardized structure including a title, learning objectives, required materials, and four key instructional sections: warm-up, main tasks (explain and reinforce), and cool-down. The focus on multiple grade levels and lesson plan sections allowed us to analyze GenAI performance across different pedagogical contexts.
    \item \textbf{Evaluation measures:}  Across pairs of lesson plans (customized GPT-4 vs. LLaMA-2-13b FT, LLaMA-2-13b FT vs. human curriculum designer, human curriculum designer vs. customized GPT-4) human raters indicated their preferred lesson plan on each of four dimensions (binary: Yes/No):
    \begin{itemize}
        \item Warm Up: Relevance, quality, and usefulness 
        \item Main Tasks (Explain \& Reinforce): Instructional effectiveness 
        \item Cool Down: Concluding quality and coherence 
        \item Overall preference: Which plan they would most likely use 
    \end{itemize}
    \item \textbf{Comments}: Participants provided written comments explaining their choices, offering deeper insights into their decision-making process.
\end{itemize}

\subsection{Study Interface}
The study was conducted via a web-based platform designed to present lesson plan pairs in a controlled, unbiased manner. Each participant was shown two anonymized lesson plans labeled as \textit{Version A} and \textit{Version B} alongside the input query details (title, grade, standard, subject) side-by-side.

\begin{figure*}[th!]
  \centering
  \includegraphics[width=0.7\linewidth]{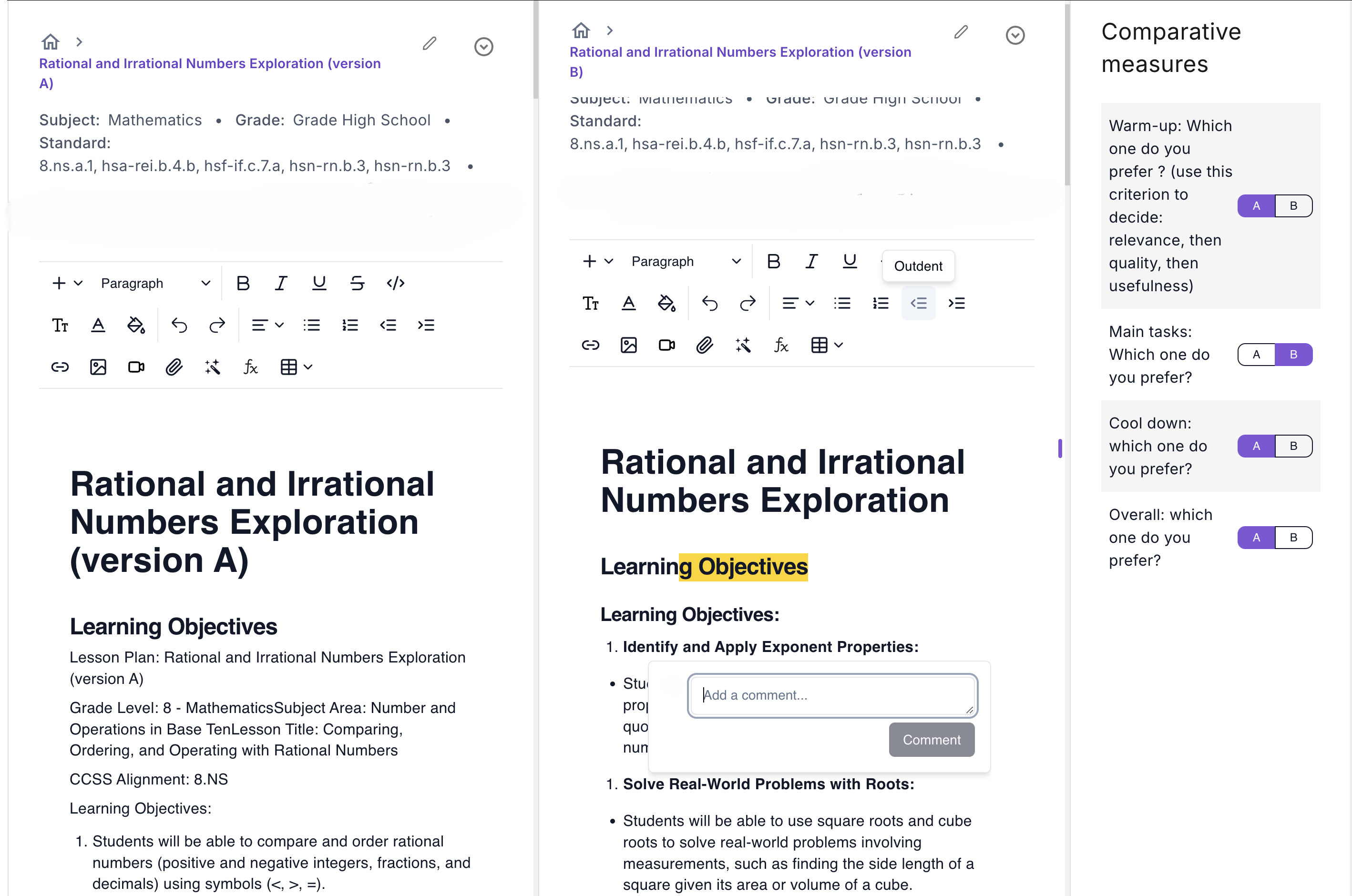}
  \caption{User study interface displaying lesson plan pairs with preference criteria and comment functionality}
  \label{fig:ui}
  \Description{}
\end{figure*}

Participants were required to carefully read both lesson plans in each pair and select which one they preferred for each measure. To prevent bias related to the authoring source, participants were blinded to the source (GenAI vs. human). They indicated preferences for each section using binary choices and were encouraged to leave qualitative comments explaining their evaluations. Figure~\ref{fig:ui} illustrates the study interface.

\subsection{Participants and Study Procedure}
Twenty experienced (with more than 10 years of teaching experience) K-12 mathematics educators and teachers participated in the study, spanning elementary, middle, and high school levels. The study followed these steps:

(i) Orientation: A virtual focus group briefing introduced participants to the study objectives and procedures. Informed consent was obtained.

(ii) Pilot training: Participants completed 2-4 practice comparisons to familiarize themselves with the evaluation criteria and interface.

(iii) Lesson plan assessments: Each participant evaluated 20 lesson plan pairs over several weeks (March-July 2024), dedicating approximately 10 hours in total (2 pairs/hour).

(iv) Weekly check-ins: Research team members assigned lesson plans tailored to participants' grade-level expertise and conducted weekly meetings to address any concerns.

\subsection{Data Collection}
We gathered both quantitative and qualitative data, including binary preferences and comments. Four participants withdrew mid-study, and their partial data were excluded to ensure consistency, resulting in a final dataset of $529$ unique evaluated lesson plan pairs.

\section{Data Analysis and Findings}

\subsection{Author Pair and Grade-Level Distribution}
Table~\ref{tab:dataset_description} provides an overview of the preferences collected in the study, detailing the distribution of lesson plan evaluations across different grade levels and author pairings. After cleaning missing and incomplete data, we obtained $529$ complete lesson plan pairs from $16$ educators, each assessed across $4$ distinct measures, resulting in $2,116$ total evaluations.

\begin{table}[ht]
  \caption{Author Pair and Grade-Level Description}
  \label{tab:dataset_description}
 %\resizebox{\textwidth}{!}{
  \begin{tabular}{l|c|c|c|c}
    \toprule
        \textbf{} & \textbf{Total} & \textbf{E} & \textbf{M} & \textbf{HS} \\
    \midrule
        \textbf{Total Lesson Pairs} & 529 & 284 & 84 & 161 \\
        \textbf{Total Measures} & 2116 & 1136 & 336 & 644 \\
    \midrule
        \textbf{Author Pair Dist.} & & & & \\
        \textbf{HCD-- CGPT-4} & 206 & 120 & 28 & 58 \\
        \textbf{HCD--LLaMA-2-13b FT} & 190 & 87 & 38 & 65 \\
        \textbf{LLaMA-2-13b FT-- CGPT-4} & 133 & 77 & 18 & 38 \\
    \midrule
        \textbf{Author Dist.} & & & & \\
        \textbf{CGPT-4} & 339 & 197 & 46 & 96 \\
        \textbf{LLaMA-2-13b FT} & 323 & 164 & 56 & 103 \\
        \textbf{HCD} & 396 & 207 & 66 & 123 \\
    \bottomrule
\end{tabular}
%}
\end{table}

The table disaggregates data by grade level -- elementary (E), middle (M), and high school (HS) -- and by author pairings -- customized GPT-4 (CGPT-4) vs. human curriculum designer (HCD), fine-tuned LLaMA-2-13b (LLaMA-2-13b FT) vs. HCD, and CGPT-4 vs. LLaMA-2-13b FT. This structure provides insights into how educator preferences varied across grade levels and lesson plan sources.

The distribution of author pairs shows a slight imbalance, with HCD–CGPT-4 (206 pairs) and HCD–LLaMA-2-13b FT (190 pairs) being relatively balanced, while CGPT-4–LLaMA-2-13b FT (133 pairs) had fewer comparisons. This imbalance was most pronounced in middle school evaluations, where CGPT-4 vs. LLaMA-2-13b FT had only 18 comparisons, limiting statistical power for that subset. In terms of lesson plan counts, HCD contributed 396 lesson plans, followed by CGPT-4 (339) and LLaMA-2-13b FT (323), suggesting a reasonably even distribution across sources.

While the overall dataset exhibits some imbalance, this does not inherently skew within-pair comparisons, as each pair is evaluated independently. Each pairwise evaluation (e.g., CGPT-4 vs. LLaMA-2-13b FT) was analyzed in its own context, ensuring that the raw pairwise preferences accurately reflect educator choices as this was the main method of the comparison in the study.

In addition to binary preference data, we collected 2,979 participant comments, offering qualitative insights into decision-making rationales. Table~\ref{tab:comment_distribution} summarizes comment distribution across author pairs and grade levels. These comments, though optional, provided critical context for understanding selection patterns. The highest engagement was observed in elementary-grade lesson plans (1,825 comments), followed by high school (646 comments) and middle school (508 comments). Author-wise, CGPT-4–HCD pairs received the most comments (1,058), while LLaMA-2-13b FT–CGPT-4 pairs also saw high engagement (968 comments). Interestingly, LLaMA-2-13b FT-HCD pairs generated slightly fewer comments (953), despite having more lesson plan pairs than the CGPT-4–LLaMA-2-13b FT group. By individual author, LLaMA-2-13b FT received the most comments (1,176), followed by CGPT-4 ($1,112$) and the HCD ($691$), indicating higher engagement, particularly at the elementary and high school levels. These suggest that educators engaged more critically with AI-generated content, particularly for elementary and high school levels.

\begin{table}[ht]
\centering
\caption{Comment Distribution in the Dataset}
\label{tab:comment_distribution}
 %\resizebox{\textwidth}{!}{
\begin{tabular}{l|c|c|c|c}
\toprule
    \textbf{} & \textbf{Total} & \textbf{E} & \textbf{M} & \textbf{HS} \\ 
\midrule
    \textbf{Total Comments} & 2979 & 1825 & 508 & 646 \\
\midrule
    \textbf{Comment Dist. by Author Pair} & & & & \\
    \textbf{HCD--CGPT-4} & 1058 & 680 & 157 & 221 \\
    \textbf{HCD--LLaMA-2-13b FT} & 953 & 479 & 235 & 239 \\
    \textbf{LLaMA-2-13b FT--CGPT-4} & 968 & 666 & 116 & 186 \\
\midrule
    \textbf{Comment Dist. by Author} & & & & \\
    \textbf{CGPT-4} & 1112 & 730 & 145 & 237 \\
    \textbf{HCD} & 691 & 419 & 145 & 127 \\
    \textbf{LLaMA-2-13b FT} & 1176 & 676 & 218 & 282 \\
\bottomrule
\end{tabular}
%}
\end{table}

Notably, $66$ lesson pairs lacked comments, as feedback was optional. While comment volume generally aligned with lesson plan distribution, variations suggest that certain author pairs elicited stronger reactions from participants, an aspect explored further in qualitative analysis.

\subsection{Analytical Methods Overview}
We employed a mixed-method approach to analyze educators' preferences, structured around the following research questions:

\textbf{RQ1:} How do educators' preferences compare across lesson plans from different authors (LLaMA-2-13b FT, CGPT-4, and HCD) in pairwise evaluations across all measures?

\textbf{RQ2:} How do educators' preferences vary across evaluation measures (warm-up, main tasks, cool-down, and overall quality)?

\textbf{RQ3:} How do preferences differ across grade levels (elementary, middle, high school)?

\textbf{RQ4:} What key factors influence educators' preferences, and how do these vary by evaluation measure and grade level?

For RQ1–RQ3, we conducted quantitative analyses using:
(i) Pairwise analysis: Direct comparisons between author pairs (CGPT-4 vs. HCD, LLaMA-2-13b FT vs. HCD, CGPT-4 vs. LLaMA-2-13b FT).
(ii) Evaluation measure breakdown: Examining how educator preferences varied across warm-up, main task, cool-down, and overall quality.
(iii) Grade-level disaggregation: Assessing whether educator preferences shifted across elementary, middle, and high school contexts.
For RQ4, we conducted qualitative analyses of participant comments using manual thematic analysis, contextual embedding, and topic modeling to extract key themes. These qualitative insights provide depth to the numerical findings, revealing the underlying rationale behind educator preferences.

\subsection{RQ1: Pairwise Comparisons Across All Measures}
To assess educator preferences for lesson plans generated by different authors, we analyzed direct pairwise comparisons across three author pairs: HCD vs. CGPT-4, HCD vs. LLaMA-2-13b FT, and CGPT-4 vs. LLaMA-2-13b FT. Each pair was evaluated separately, with participants making binary preferences (Y/N) for each pair, ensuring that the results directly reflect actual decision-making in the study.

\begin{figure}[htbp]
        \centering
        \includegraphics[width=\linewidth, keepaspectratio]{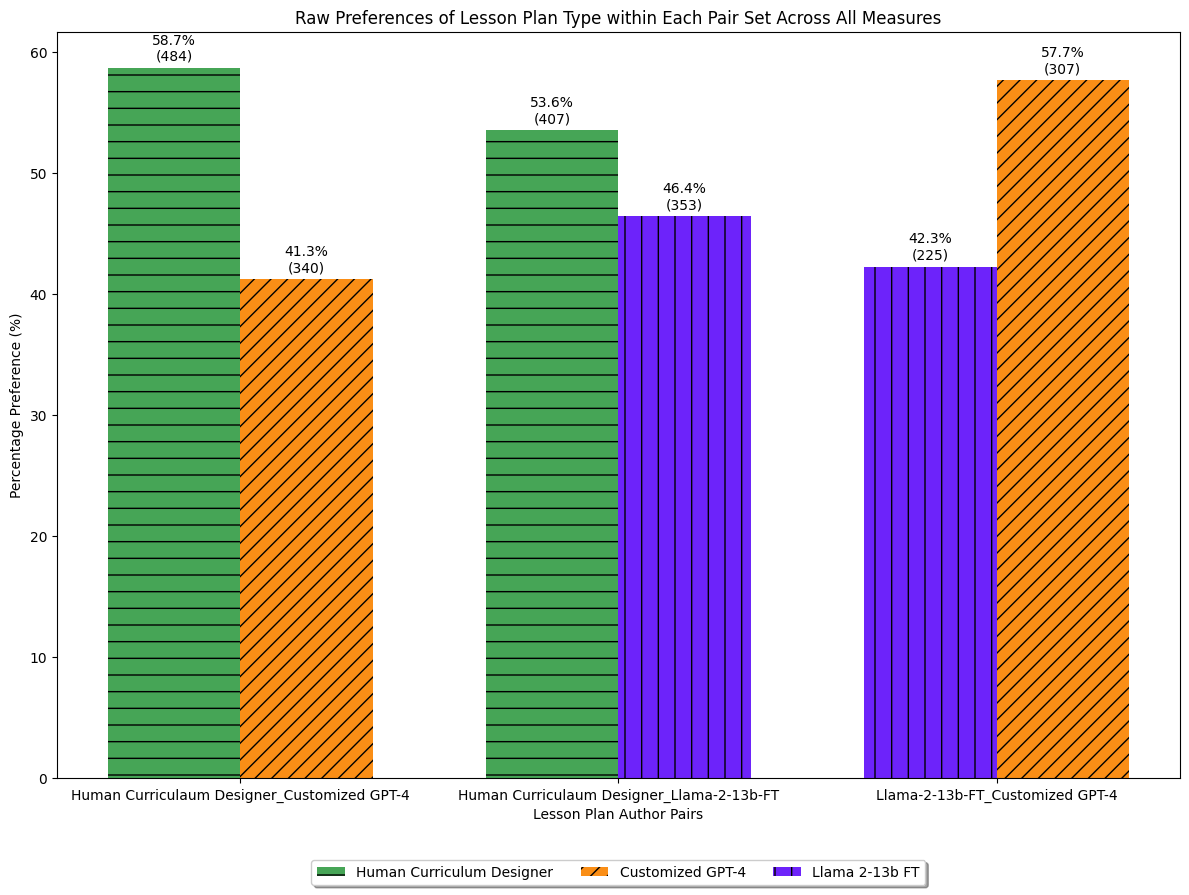}
        \caption{Educators' preferences for individual lesson plan type across all measures within each author-pair}
        \label{fig:rq1_pair_raw}
\end{figure}

Figure~\ref{fig:rq1_pair_raw} presents the raw pairwise preferences across evaluation measures. CGPT-4 was preferred 57.7\% of the time, while LLaMA-2-13b FT was preferred 42.3\% of the time -- a 15.4\% margin favoring CGPT-4. This suggests that prompt-engineered CGPT-4 lesson plans better aligned with educator expectations, particularly in content structuring and instructional coherence. Educators preferred HCD-created lesson plans 58.7\% of the time, while CGPT-4 received 41.3\%. This 17.4\% preference gap suggests that while AI-generated plans are gaining acceptance, human-created plans still better meet pedagogical needs. Lesson plans by HCD were preferred 53.6\% of the time, while LLaMA-2-13b FT received 46.4\% of preferences. The relatively small (7.2\%) margin suggests that fine-tuned LLaMA-2-13b produced competitive lesson plans, albeit slightly less favored than human-created content. These direct pairwise comparisons provide the most meaningful insight into educator decision-making, as they reflect actual choices made in the study.

\subsection{RQ2: Preferences by Specific Evaluation Measures}
To address RQ2, we analyzed educator preferences across four key lesson plan evaluation measures -- warm-up, main tasks, cool down, and overall quality -- to determine which lesson plan sections different authors excel in and where improvements are needed. A preference for a particular measure for an lesson plan author suggests that educators found that section of the lesson plan better aligned with their instructional goals and expectations.

\begin{figure*}[htbp]
     \centering
        \includegraphics[width=0.90\linewidth, keepaspectratio]{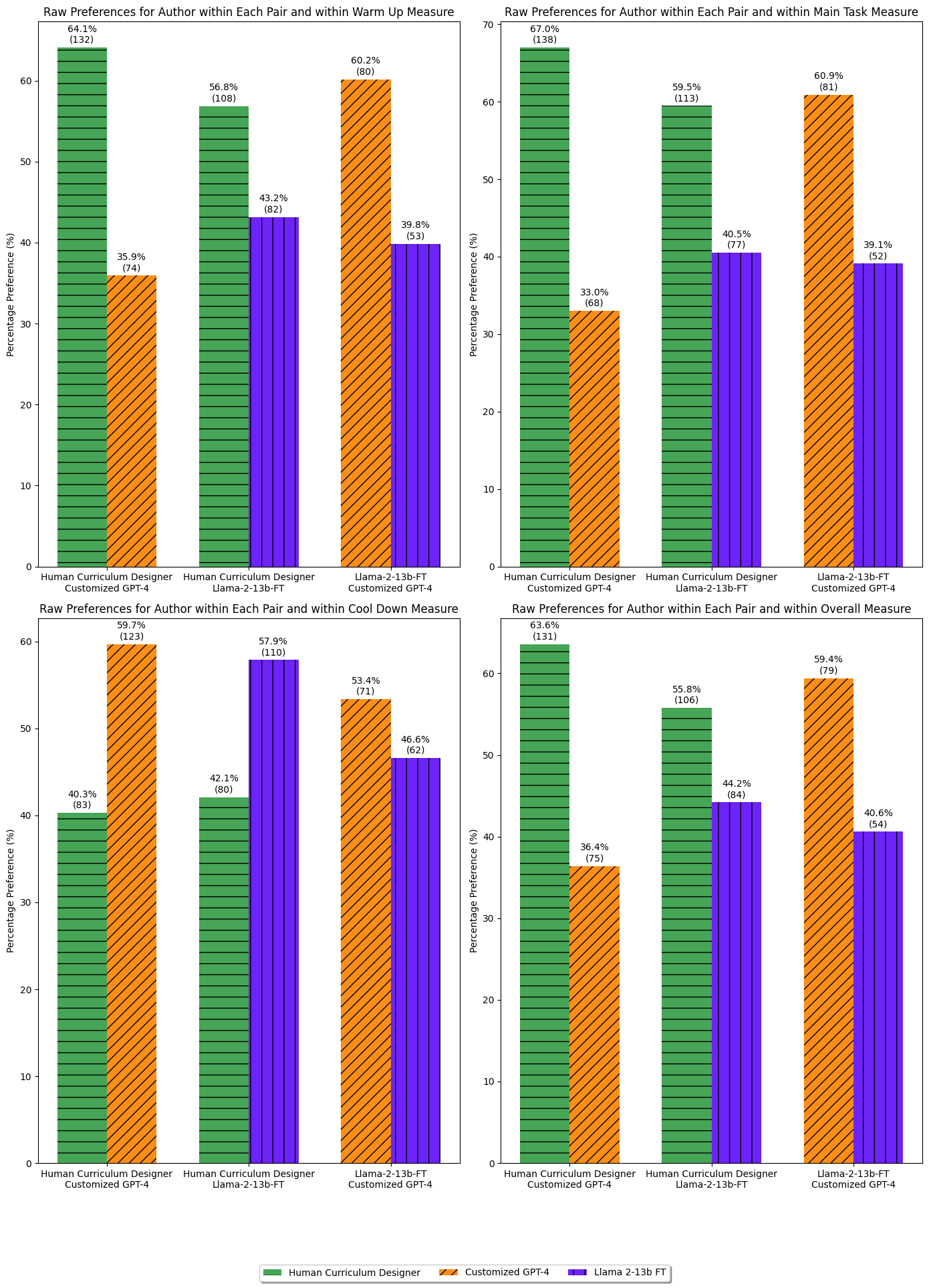}
        \caption{Educators' preferences for individual lesson plan type within each measure and within each author-pair}
        \label{fig:rq2_pair_raw}
\end{figure*}

Pairwise comparisons (Figure~\ref{fig:rq2_pair_raw}) provide direct insight into how preferences shifted between specific lesson plan sections. 

For warm-up, HCDs were preferred over both GenAI models, with 64.1\% over CGPT-4 and 56.8\% over LLaMA-2-13b FT. Between the GenAI models, CGPT-4 (60.2\%) outperformed LLaMA-2-13b FT (39.8\%), indicating that while AI-generated warm-up activities are improving, educators still trust human-designed introductions more. 

For main tasks, HCDs again dominated, preferred 67.0\% over CGPT-4 and 59.5\% over LLaMA-2-13b FT. Between the GenAI models, CGPT-4 (60.9\%) outperformed LLaMA-2-13b FT (39.1\%), highlighting its advantage in generating instructional content.

The cool-down section revealed a notable shift in preference toward GenAI models. CGPT-4 (59.7\%) was preferred over HCD (40.3\%), and LLaMA-2-13b FT (57.9\%) also outperformed human-created lesson conclusions (42.1\%). This suggests that GenAI-generated lesson wrap-ups are particularly effective, likely due to their ability to generate structured, reflective summaries.

For overall lesson plan quality, HCDs were still preferred 63.6\% over CGPT-4 and 55.8\% over LLaMA-2-13b FT, though GenAI-generated content showed increasing acceptance. These findings confirm that GenAI-generated lesson plans are gaining traction, particularly in certain sections like cool-down activities, while human-created plans remain the preferred choice for structured instructional tasks.

To further validate these pairwise comparisons and account for dataset imbalances, we conducted a secondary analysis over the aggregated preferences within each evaluation measure across all author comparisons to assess the robustness of observed preferences. To mitigate the dataset imbalances as mentioned above, we did this analysis as a bootstrapping analysis (1,000 resamples, 95\% confidence intervals, standard error estimation)

\begin{figure*}[htbp]
        \centering
        \includegraphics[width=\linewidth, keepaspectratio]{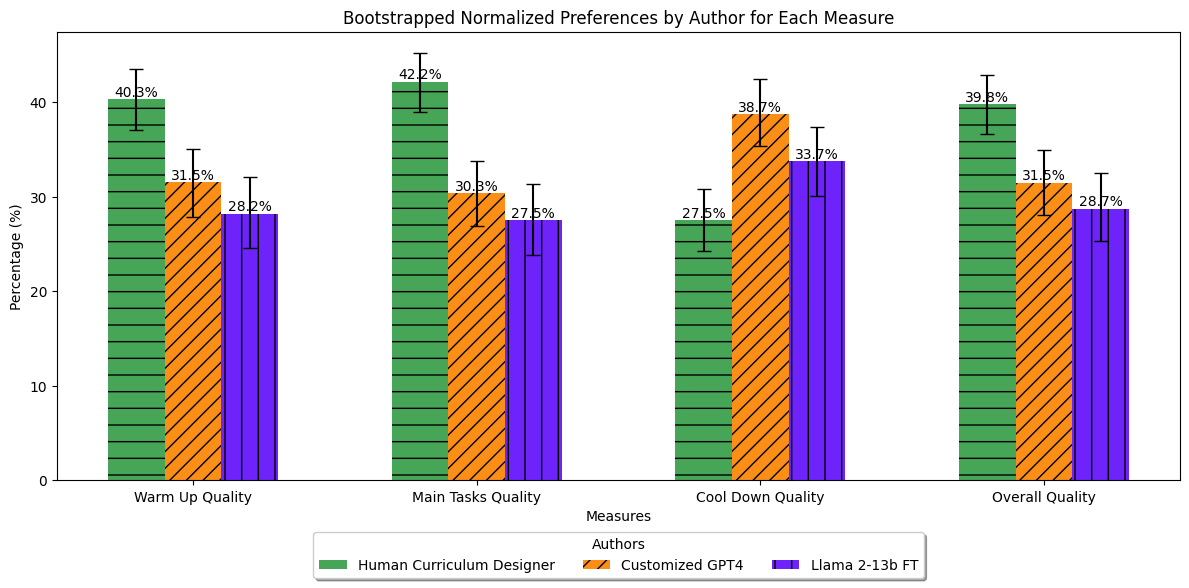}
        \caption{Educators' preferences for individual lesson plan type across all measures (Robust)}
        \label{fig:rq2_all_bootstrap}
\end{figure*}

Bootstrapping analysis (Figure~\ref{fig:rq2_all_bootstrap}) reaffirmed HCD as the most preferred for warm-up (40\%), main tasks (42.2\%), and overall quality (39\%). For cool-down activities, CGPT-4 (38\%) had a clear lead, confirming its effectiveness in structuring lesson wrap-ups. LLaMA-2-13b FT also performed well in cool-down activities, scoring 33.7\%, surpassing human-designed plans, reinforcing that GenAI-generated content may be particularly well-suited for summarization and reinforcement activities. Notably, the confidence intervals for GenAI-generated lesson plans were narrower in cool-down measures, indicating greater consistency in educator evaluations. This suggests that while GenAI models generally received lower average ratings, their assessments were less variable than those for human-generated content.

\subsection{RQ3: Preferences Across Grade Levels}
To examine RQ3, we analyzed educator preferences across elementary (grades K-5), middle (grades 6-8), and high school (grades 9-12), exploring how perceptions of different lesson plan types varied with student age and curriculum complexity.

Pairwise comparisons (Figure~\ref{fig:rq3_pair_raw}) reveal distinct trends across grade levels:

\begin{figure*}[htbp]
     \centering
        \includegraphics[width=\linewidth, keepaspectratio]{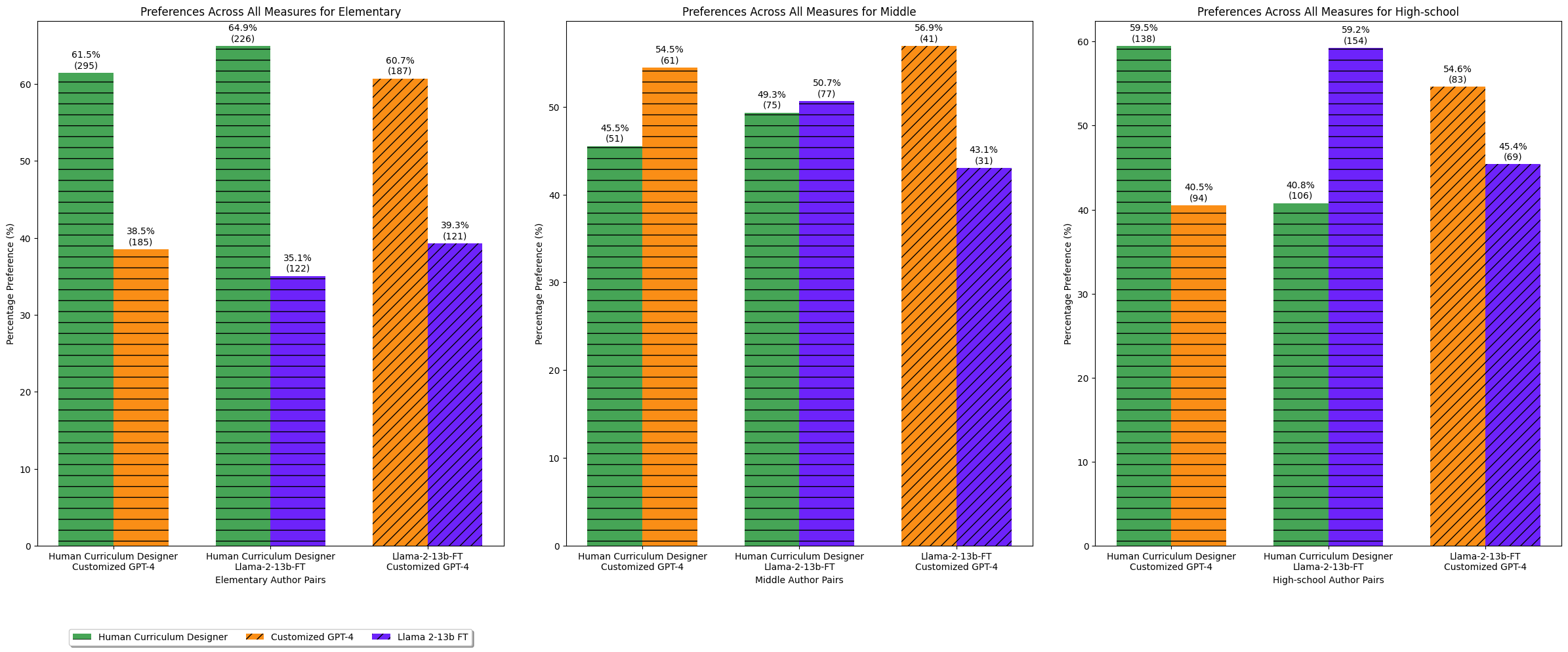}
        \caption{Educators' preferences for individual lesson plan type across all measures within each author-pair}
        \label{fig:rq3_pair_raw}
\end{figure*}

In elementary school, HCD-created lesson plans were overwhelmingly preferred across all measures. HCDs were chosen 61.5\% over CGPT-4 and 64.9\% over LLaMA-2-13b FT, reaffirming that educators rely more on human-authored lesson plans for younger students, likely due to their structured engagement and developmental alignment. Between GenAI models, CGPT-4 outperformed LLaMA-2-13b FT in 60.7\% of comparisons, suggesting that a well-prompted, customized GPT-4 generates more educator-aligned content than a fine-tuned LLaMA-2-13b model at this level.

In middle school, educator preferences became more competitive. CGPT-4 leads in most comparisons. CGPT-4 was preferred 54.5\% over HCD (45.5\%) and LLaMA-2-13b FT (50.7\%) was preferred over HCD (49.3\%) suggesting that educators begin to see AI-generated lesson plans as more viable alternatives. Interestingly, CGPT-4 (56.9\%) was favored over LLaMA-2-13b FT (43.1\%), maintaining its lead among AI-generated content. This indicates that for middle school instruction, structured and customized AI outputs start competing effectively with human-designed content.

In high-school, the preference landscape also shifts. LLaMA-2-13b FT outperforms HCD significantly marking a critical transition where a fine-tuned AI model becomes more effective in generating high school content. Educators preferred LLaMA-2-13b FT 59.2\% over HCD (40.8\%), suggesting that domain-specific fine-tuning helps the model better capture high school-level complexity. However, CGPT-4 still led in direct comparisons with LLaMA-2-13b FT, with a preference margin of 54.6\% over 45.4\%, showing that its structured content generation was well-received.

These findings highlight a clear grade-dependent transition in AI-generated lesson plan acceptance. While human-created lesson plans remain the gold standard in elementary education, high school shows a more competitive preference landscape, and AI models gain an edge in middle school, likely due to their ability to handle complex subject matter.

\subsection{RQ4: Key Reasons Behind Preferences for Lesson Plan Types}
To address RQ4, which examines the rationale behind educators' preferences for different lesson plan types, we employed two complementary qualitative analysis techniques: Latent Dirichlet Allocation (LDA) topic modeling and manual thematic analysis. LDA was used to extract frequently discussed themes from educator comments, offering a machine-learning-driven method to identify patterns in qualitative feedback. However, LDA does not distinguish whether a particular theme was discussed positively or negatively, nor does it indicate whether a comment was made in favor of or against a lesson plan. To address this limitation, a research team member performed manual thematic analysis, reading and categorizing all comments to determine whether specific themes were associated with reasons for selecting or rejecting a lesson plan. This step allowed for a deeper contextual interpretation that LDA alone could not provide.

To determine the optimal number of topics for LDA, we used coherence scores, perplexity, lower bound, and exclusivity measures to ensure that extracted topics were both meaningful and distinct. After identifying the final topic set, we aligned them with established educational constructs to define coherent thematic categories. These categories were further refined by incorporating findings from the manual thematic analysis, ensuring that the topics reflected both machine-detected patterns and researcher-validated interpretations.

While only one research team member conducted the thematic analysis, limiting inter-coder reliability, this approach was deemed appropriate due to the concise nature of educator comments. Unlike longer qualitative responses that require multiple coders to ensure interpretive consistency, the comments in this study were short and relatively unambiguous, reducing the likelihood of subjective bias in thematic assignment. Additionally, the consistency between LDA-extracted themes and researcher-identified themes supports the robustness of the analysis.

\begin{figure}[htbp]
    \centering
        \centering
        \includegraphics[width=\linewidth]{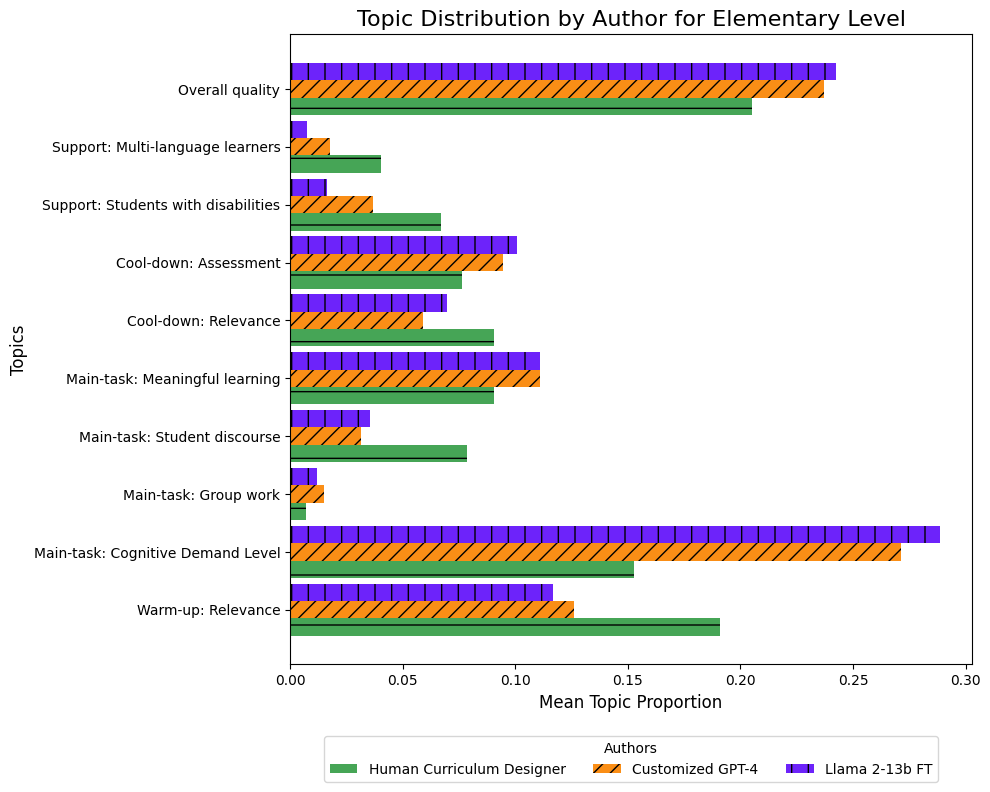}
        \caption{Topic Distribution by Author for Elementary Level: Thematic analysis of educator comments categorized by key instructional elements, showing variations in emphasis across HCD, CGPT-4 LLaMA-2-13b FT.}\label{fig:topic_elementary}
\end{figure}

\begin{figure}[htbp]
        \centering
        \includegraphics[width=\linewidth]{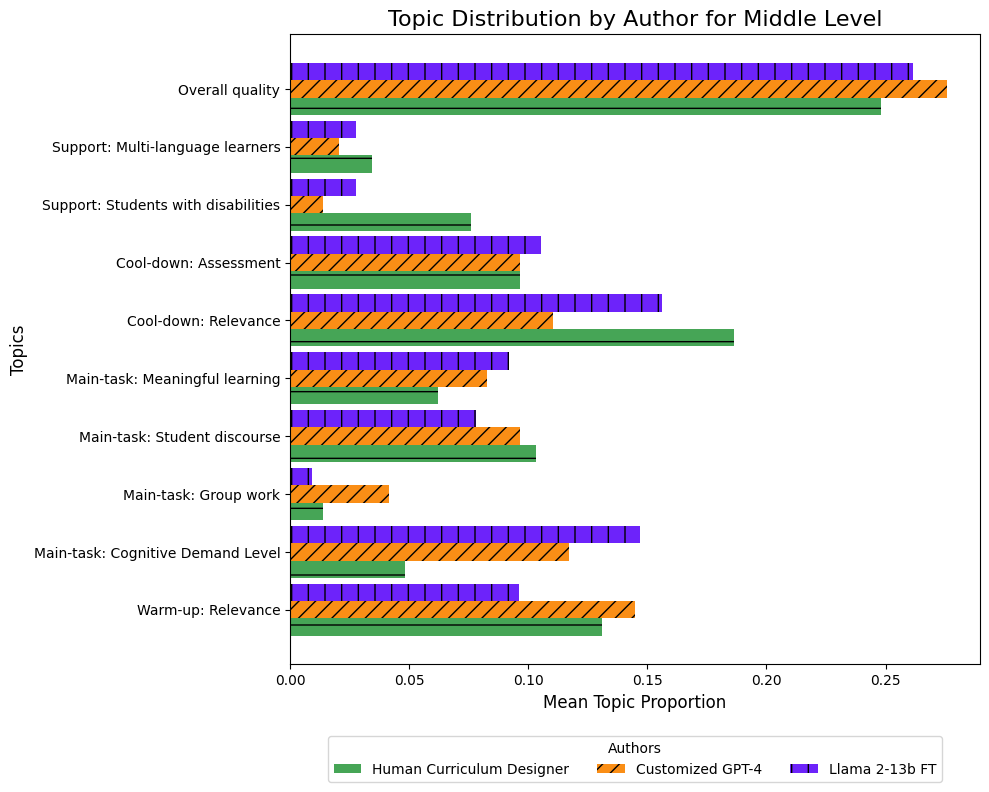}
        \caption{Topic Distribution by Author for Middle School Level}\label{fig:topic_middle}
\end{figure}

\begin{figure}[htbp]
        \centering
        \includegraphics[width=\linewidth]{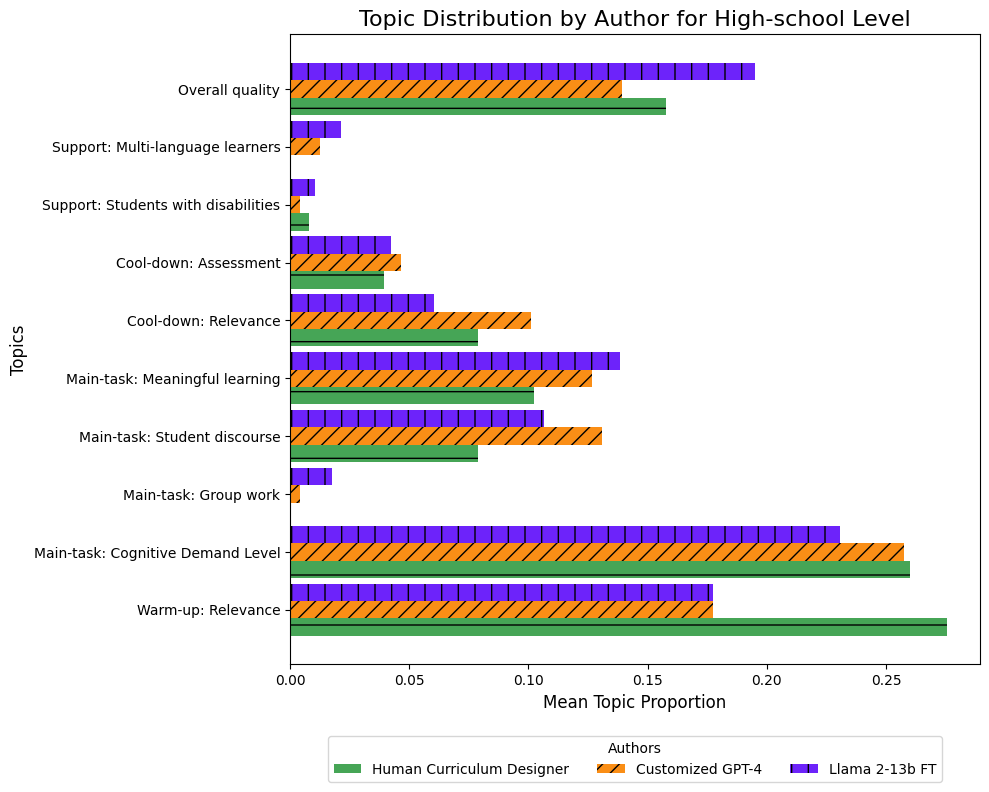}
        \caption{Topic Distribution by Author for High School Level}\label{fig:topic_high}
\end{figure}

Figures~\ref{fig:topic_elementary}, \ref{fig:topic_middle}, and \ref{fig:topic_high} present the key themes identified across elementary, middle, and high school grade levels, highlighting how different lesson plan elements influenced educator preferences. These themes include warm-up relevance, cognitive demand level, group work, meaningful student discourse, and supportive strategies for students with disabilities and multi-language learners. 

LDA topic modeling revealed several recurring themes and our manual examination of the comments supplement the content context.

As we see, cool-down assessment was a moderate discussion point across all models, with CGPT-4 and LLaMA-2-13b FT receiving more feedback than HCDs. Educators valued CGPT-4's structured approach to assessments, reinforcing RQ2 findings that AI-generated lesson conclusions were well-received. One participant mentioned: \textit{"Strong cool down with problem solving and student discourse and reflection"}.

Group work was another key theme. While CGPT-4 and LLaMA-2-13b FT received more comments, educators mainly preferring HCD-authored lesson plans for fostering structured collaboration, particularly in elementary and middle school. CGPT-4 showed some ability to incorporate group activities, but educators noted a lack of depth. 

Educators also frequently commented on the cognitive demand level of lesson plans. LLaMA-2-13b FT was highly praised for its rigorous, structured content in high school instruction, reinforcing RQ3 findings that AI-generated lesson plans are more competitive at higher grade levels. However, HCD-authored plans were preferred for balancing rigor with engagement, especially for younger students, suggesting that GenAI-generated content may require human adaptation for effective classroom use.

Student discourse emerged as a major distinction between AI-generated and human-authored lesson plans. HCD lesson plans were strongly preferred for fostering meaningful discussions, with educators noting that GenAI-generated content often lacked explicit prompts for student interactions. This was most evident in middle school, where structured discourse is critical for conceptual learning. For instance, one participant mentioned: \textit{"The main task in this version is more aligned with the standards, activities move from simpler to more complex tasks and involves student interactions."}

Warm-up relevance showed grade-dependent trends. HCDs were highly rated for structured warm-ups in high school and elementary reflecting the importance of developmentally appropriate engagement strategies that GenAI models still struggle to replicate. 

Support for students was another important theme, particularly for multi-language learners and students with disabilities. HCD lesson plans received the most positive feedback for supportive strategies, with educators highlighting scaffolded supports and differentiated instruction. GenAI-generated plans were often described as "neutral" in this regard, meaning they lacked targeted support rather than actively excluding students.

Overall lesson quality was the most general theme, with CGPT-4 receiving positive feedback for structured content delivery, especially in high school. However, across all grade levels, HCD-authored lesson plans remained the preferred choice due to their adaptability, pedagogical soundness, and ability to engage students. This aligns with RQ1 and RQ2 findings that human-authored content still sets the standard. Educators' comments suggest that while GenAI is competitive, particularly at higher grade levels, HCD-authored content remains the gold standard in terms of quality and effectiveness.

\subsection{RQ1-RQ4: Key Insights from the Data Analysis}
Findings from this analysis highlight how GenAI-generated lesson plans are being integrated into K-12 education, with varying degrees of acceptance based on grade level.

(i) GenAI models are increasingly competitive at higher grade levels. In middle school, GenAI models -- particularly CGPT-4 -- become more competitive, with educators showing a growing openness to AI-generated content. In high and middle schools, LLaMA-2-13b FT surpasses HCDs in many cases, indicating that fine-tuned AI models can effectively handle complex, subject-specific content. This suggests that fine-tuned models may be better suited for handling advanced curriculum complexity compared to general-purpose AI models. RQ 4 findings showed that LLaMA-2-13b FT excelled in content depth, aligning with RQ3's finding that AI-generated plans are more competitive in upper-grade instruction. 

(ii) CGPT-4 consistently excels in cool-down activities. Across all grade levels, it was preferred for lesson conclusions, suggesting that GenAI models are effective in reinforcing learning and guiding students through structured reflections.

(iii) HCD-created lesson plans remain dominant in elementary education, where structured engagement and developmental appropriateness are critical.

\section{Discussion}
The findings from this study highlight the evolving role of GenAI in lesson planning, revealing both its strengths and persistent challenges. Educators demonstrated a growing openness to AI-generated lesson plans, particularly in structured instructional components such as cool-down assessments. However, human curriculum designers (HCD) remained the gold standard, particularly in fostering student discourse, group work, and inclusive instructional strategies. These results indicate the need for a more refined approach to GenAI-assisted lesson plan design, where GenAI models are optimized based on direct educator feedback. Refining GenAI models to better align with teacher preferences requires incorporating user engagement data into model training, emphasizing elements such as scaffolded support, differentiated instruction, and interactive learning strategies. AI models, particularly LLaMA-2-13b FT, demonstrated potential for high cognitive demand tasks at the high school level, suggesting that fine-tuning AI models for subject-specific content could enhance their effectiveness in more advanced educational settings.

These findings also call for a redefinition of evaluation measures in AI-generated lesson plans. While traditional metrics such as content accuracy and coherence remain important, educators placed significant emphasis on lesson adaptability, student support, and the facilitation of student engagement -- areas where AI models require further refinement. The study highlights the importance of integrating AI with professional curriculum design expertise rather than viewing the two as competing approaches. Therefore, best practices for GenAI-human collaboration should focus on leveraging AI's efficiency in content structuring while maintaining the nuanced pedagogical insights that human educators bring to lesson planning.

\section{Assumptions and Limitations}
As with all other research studies, our study is not free of certain limitations and assumptions that may influence our findings.

Given the study's reliance on qualitative feedback and subjective preferences, we assume that participants' responses accurately reflect their instructional needs and experiences in evaluating lesson plans at the component level (e.g., warm-up, main tasks, cool-down). We also assume that pairwise comparisons provide meaningful insights into educator preferences, as each evaluation was conducted independently. While we aimed for a diverse educator sample, individual teaching philosophies, familiarity with AI tools, and regional pedagogical norms may influence results, limiting the generalizability of our findings.

While this study presents valuable insights, its generalizability is limited by the geographic, demographic, and contextual diversity of the educator sample. Pedagogical priorities may differ across regions, school types, and educational systems, which could influence educator preferences for AI-generated content. Additionally, educator biases, familiarity with AI tools, and subjective experiences may have shaped their evaluations, making it difficult to derive universally applicable conclusions. Additionally, while the study focused on K-12 mathematics, findings may not directly apply to other subjects that emphasize different instructional strategies.

In future, we aim to expand our participant diversity across regions, subjects, and teaching styles to ensure lesson plans are evaluated in a broad range of educational contexts. We will conduct longitudinal studies to track AI's long-term impact on student learning outcomes and educator workload, providing deeper insights into its practical applications. We also plan to assess how GenAI influences educator engagement, student motivation, and learning retention through lesson planning, ensuring that GenAI-created lesson plans are practical, ethically responsible, and aligned with real-world classroom needs.

\section{Conclusion}
This study reinforces GenAI's role as a collaborative tool rather than a replacement for educator expertise. By generating structured, adaptable lesson plans, GenAI can reduce educator' cognitive load, allowing them to focus on tailoring content to diverse student needs. Through comparative analysis, we evaluated Customized GPT-4 and fine-tuned LLaMA-2-13b against human curriculum designer-created lesson plans, highlighting both AI's strengths in structured tasks and its limitations in deep pedagogical reasoning.

The findings suggest that fine-tuning GenAI with domain-specific datasets and incorporating education feedback loops can significantly enhance AI-generated lesson plans. While human expertise remains essential for complex instructional design, GenAI shows potential as an efficient planning assistant, particularly for structured learning activities and lesson summarization.

Thus by systematically evaluating AI-generated versus human-created lesson plans, this study contributes empirical evidence to the ongoing discourse on AI's role in education. AI should empower educators, fostering a balanced, collaborative approach that enhances learning outcomes while preserving the creativity and adaptability that define effective teaching.

\begin{acks}
This research was supported by the Institute of Education Sciences (IES), U.S. Department of Education, through grant \#R305C240012, as well as multiple awards from the National Science Foundation (\#2043613, \#2300291, \#2405110), and an NSF SBIR/STTR award to Hensun Innovation LLC (\#2423365). The views expressed are solely those of the authors and do not necessarily reflect the opinions of the funding agencies.
\end{acks}
\clearpage
\bibliographystyle{ACM-Reference-Format}
\bibliography{sample-base}

\end{document}